\documentclass[sigconf]{acmart}

\renewcommand\footnotetextcopyrightpermission[1]{} 

\usepackage{amsmath}
\usepackage{amsfonts}
\usepackage{graphicx}
\usepackage{textcomp}
\usepackage{xcolor}
\usepackage{multirow}
\usepackage{soul}
\usepackage{tikz}
\usepackage{algorithm}
\usepackage{algpseudocode}
\usepackage{colortbl}
\usepackage{bm}
\usepackage{float}
\usepackage{filecontents}

\newcommand{\cmmnt}[1]{}  

\def\BibTeX{{\rm B\kern-.05em{\sc i\kern-.025em b}\kern-.08em
    T\kern-.1667em\lower.7ex\hbox{E}\kern-.125emX}}

\newcommand\encircle[1]{%
\tikz[baseline=(X.base)] 
  \node (X) [draw, scale=0.75, shape=circle, inner sep=0, fill=black, text=white, minimum size=0em] {\strut #1};}

\usepackage{pifont}

\begin{document}

\title{Lightator: An Optical Near-Sensor Accelerator with Compressive Acquisition Enabling Versatile Image Processing\vspace{-0.5em}}
\author{Mehrdad Morsali$^\dagger$, Brendan Reidy$^{\ddagger}$, Deniz Najafi$^\dagger$, Sepehr Tabrizchi$^\S$, Mohsen Imani$^*$,\\ Mahdi Nikdast$^{**}$, Arman Roohi$^\S$, Ramtin Zand$^{\ddagger}$, and Shaahin Angizi$^\dagger$ \vspace{0.3em}\\
\small $^\dagger$ Department of Electrical and Computer Engineering, New Jersey Institute of Technology, Newark, NJ, USA\\
$^\ddagger$ Department of Computer Science and Engineering, University of South Carolina, Columbia, SC, USA\\
$^\S$ School of Computing, University of Nebraska–Lincoln, Lincoln, NE, USA\\
$^*$ Department of Computer Science, University of California Irvine, Irvine, CA, USA\\ 
$^{**}$ Department of Electrical and Computer Engineering, Colorado State University, Fort Collins, CO, USA\\
m.imani@uci.edu, mahdi.nikdast@colostate.edu, aroohi@unl.edu, ramtin@cse.sc.edu, shaahin.angizi@njit.edu \vspace{-1em}
\\}
\vspace{1.5em}

\begin{abstract}
This paper proposes a high-performance and energy-efficient optical near-sensor accelerator for vision applications, called Lightator. Harnessing the promising efficiency offered by photonic devices, Lightator features innovative compressive acquisition of input frames and fine-grained convolution operations for low-power and versatile image processing at the edge for the first time. This will substantially diminish the energy consumption and latency of conversion, transmission, and processing within the established cloud-centric architecture as well as recently designed edge accelerators. Our device-to-architecture simulation results show that with favorable accuracy, Lightator achieves 84.4 Kilo FPS/W and reduces power consumption by a factor of $\sim$24$\times$ and 73$\times$ on average compared with existing photonic accelerators and GPU baseline.\vspace{-1.7em}
\end{abstract}




\maketitle
\pagestyle{plain} 

\section{Introduction}
While the prevalence of the Internet of Things (IoT) has grown significantly, it still lacks inherent intelligence and heavily depends on cloud-based decision-making. In such a cloud-oriented paradigm, a considerable portion of data created by IoT sensors remains unprocessed \cite{song2022reconfigurable,xu2020macsen}. Vision sensors typically capture light and convert it into electrical signals, which are subsequently stored, processed, transmitted, and utilized. This procedure necessitates the transformation of all individual pixels into predetermined digital values with a fixed bit-width (e.g., 8 bits \cite{xu2020macsen,angizi2023pisa}). Remarkably, the major share of power consumption in traditional vision sensors, exceeding 96\% \cite{xu2020macsen}, is ascribed to the conversion and retention of pixel values. This is predominantly associated with memory- and computation-intensive algorithms and the limited processing capabilities of current IoT devices, which are constrained by power and size limits \cite{tang2019considerations,angizi2023pisa}. To confront these challenges, a shift from a cloud-oriented to a thing-centered (data-centric) approach is imperative, wherein IoT nodes locally process the data \cite{hsu2019ai}.

Recent efforts have focused on enhancing CMOS image sensors for faster processing of Deep Neural Network (DNN) workloads. One approach is the integration of CMOS image sensors and processors on a single chip, known as Processing-Near-Sensor (PNS) \cite{carey2013100,hsu20200,yamazaki20174}. Another method involves incorporating computation units with individual pixels, termed Processing-In-Sensor (PIS) \cite{xu2020macsen,xu2021senputing,tabrizchi2023appcip,angizi2023pisa}. The PIS platform processes pre-Analog-to-Digital Converter (pre-ADC) data before transmitting it to the on-/off-chip processor. Despite these advancements, some challenges persist, including the energy consumption of ADC, Digital-to-Analog Converters (DAC), and sense amplifiers in PIS, limiting the deployment of all DNN layers into the pixel array \cite{el1999pixel,song2022reconfigurable}. 

Most studies have focused on accelerating the initial layer and outsourcing the remaining layers to a digital accelerator due to the restricted resources of PIS. Therefore, three key challenges remain unaddressed in current electronic PIS/PNS designs: $(i)$ power-hungry peripherals and ADC/DAC units, even when reduced for sensing and computing \cite{choi2015energy,xu2020macsen,hsu2019ai,sunny2021crosslight}; $(ii)$ significant area overhead and power consumption in recent PNS/PIS units, necessitating additional memory for intermediate data storage \cite{angizi2023pisa,tabrizchi2023appcip,song2022reconfigurable}; and $(iii)$ constrained computation speed due to electronic systems operating at a few GHz, lacking the capability to support the high speeds and extensive parallelism observed in optical systems with photo-detection rates exceeding 100GHz \cite{sunny2021robin,sunny2021crosslight,cheng2020silicon}.

With further advancement of integrated photonic devices (e.g., energy-efficient and tunable Microring Resonators (MRs) and Mach-Zehnder modulators), CMOS-compatible silicon photonics have emerged as a promising and viable alternative to digital electronics for building high-speed and energy-efficient optical DNN accelerators, as evidenced by several research studies \cite{shiflett2021albireo,sunny2021crosslight,liu2019holylight,zokaee2020lightbulb,sunny2021robin}, though the edge deployment of such MR devices has been insufficiently explored. Besides, even the existing MR-based accelerators have faced several challenges that this work aims to solve including $(i)$ excessive use and tuning power overhead of MRs in accelerators for activation parameters \cite{sunny2021crosslight,sunny2021robin}; $(ii)$ high power and area overhead resulting from excessive using of ADCs and DACs \cite{liu2019holylight,shiflett2021albireo,sunny2022silicon}; $(iii)$ limited flexibility in processing various DNN layers (Pooling, etc.) with no compression support; and $(iv)$ lack of correlated hardware mapping methodologies to support various kernel sizes in DNNs.
The key contributions of this work are as follows: (1) we propose a high-performance and energy-efficient optical PNS accelerator for vision applications called Lightator that can fully process various DNN layers with weight-based optical cores without relying on the cloud; (2) to meet the physical limitation of the photonic domain and power budget of IoT devices, we create innovative microarchitectural and circuit-level strategies for Lightator that enables compressive acquisition of input frames and novel hardware partitioning and mapping mechanisms to support various DNN kernel sizes; (3) we establish a solid device-to-architecture evaluation framework from the ground up and conduct thorough performance analysis and comparison of our proposed designs with state-of-the-art optical and electronic accelerator designs.
\vspace{-1.1em}

\section{Background and Related Work}
\begin{figure}[t] 
\centering
\includegraphics [width=0.78\linewidth,]{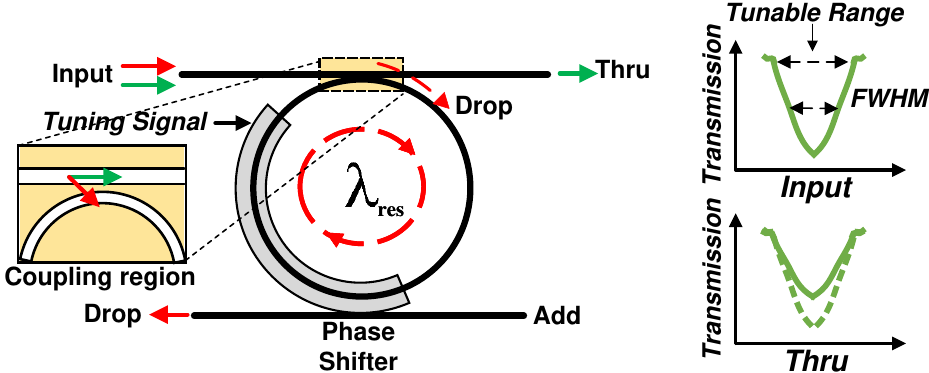}
\vspace{-1.2em}
\caption{MR input and through ports’ spectra after imprinting a parameter (using tuning signal). By adjusting the MR's resonant wavelength ($\lambda_{res}$) using the phase shifter, part of the input signal drops into the ring (through the coupling region) towards the drop port while the remaining propagates towards the through port, hence imprinting any parameter in the transmitted signals. FMHW is the full width at half maximum of the resonance spectrum.} 
\vspace{-2.3em}
\label{mr}
\end{figure}

Offering notably elevated operational bandwidth compared to electronic accelerators along with addressing fan-in/fan-out problems make silicon-photonic-based accelerators a promising candidate to accelerate DNN and machine vision applications \cite{sunny2021arxon,sunny2021crosslight,liu2019holylight,zokaee2020lightbulb}. Such accelerators can be broadly categorized into two primary designs: \textit{coherent} and \textit{non-coherent} architectures. Within the coherent category, a single wavelength is employed for operations, and weight/activation parameters are incorporated into the electrical field amplitude, phase, or polarization of an optical signal \cite{zhao2019hardware}.  Conversely, the non-coherent designs \cite{sunny2021robin,sunny2021crosslight} employ multiple wavelengths each of which capable of conducting computations concurrently. Within non-coherent architectures, the weight and input parameters of DNN are imprinted upon the signal's amplitude \cite{sunny2021crosslight,sunny2021robin}. To manipulate individual wavelengths, MRs---depicted in Fig. \ref{mr}---can be employed whose central frequency can be actively adjusted (i.e., through tuning mechanisms using, e.g., microheaters or PIN junctions), to selectively interact with specific wavelengths. 
By appropriately tuning the MRs, the incoming light intensity of a specific wavelength can be weighted. In non-coherent designs \cite{sunny2021crosslight,sunny2021robin}, MRs as a fundamental component hold the weight and activation values to be utilized in the Multiply-and-ACcumulate (MAC) operation. During photonic MAC, the transmission spectrum of input lights can be multiplied by the value adjusted on the MRs (through applying a tuning signal, see Fig. \ref{mr}). Such a value is adjusted by tuning the resonant wavelength of the MR which can partially overlap with the wavelength of the input signal, to imprint the parameter into the transmission spectrum of the input signal (see Fig. \ref{mr}). The resonant wavelength is given by $\lambda_{res}=\frac{{n_{eff}}\times L}{m}$, where $n_{eff}$ is the effective refractive index of the MR, and $L$ and $m$ denote MR’s circumference and order of the resonant mode. \cite{bogaerts2012silicon}.

Previous studies have explored accelerating DNNs through the application of both coherent and non-coherent photonic principles. LightBulb \cite{zokaee2020lightbulb} as a fully binarized Convolutional Neural Network (CNN) accelerator has been proposed which replaces the floating-point MAC operations with photonic XNOR and popcounts. With reduced computation latency and memory storage, LightBulb's excessive ADCs increased the power consumption of the design. Robin \cite{sunny2021robin} also presents an MR-based binary CNN accelerator, optimizing electro-optic components across device, circuit, and architecture layers. Despite circuit-level tuning enhancements to reduce inference latency, the excessive number of MRs and subsequent DACs required for the tuning process reduced the efficiency of the design. CrossLight \cite{sunny2021crosslight} as a 4-bit weight-input CNN accelerator requires tuning both activation and weight values in the MRs and only supports convolution layer processing similar to the previous designs. The design in \cite{sunny2022silicon} proposes a CNN accelerator with mixed-precision weight-input support. This non-coherent silicon photonic accelerator utilizes both Wavelength-Division Multiplexing (WDM) and Time-Division Multiplexing (TDM). However, the persistent use of DACs and ADCs as inter-layer transformers is a notable concern which increased the overall area and power consumption of the entire architecture. HolyLight \cite{liu2019holylight} as a nanophotonic accelerator enhances the inference throughput of CNN by using MR-based adders and shifters instead of ADCs. Nevertheless, over-utilization of MRs for both activation and weight values not only increased overall delay and power consumption but also reduced its flexibility to be used for various DNNs.  
\vspace{-0.8em}

\section{Lightator Architecture}
\begin{figure}[b]
\begin{center}\vspace{-2.5em}
\begin{tabular}{c}
\includegraphics [width=0.96\linewidth]{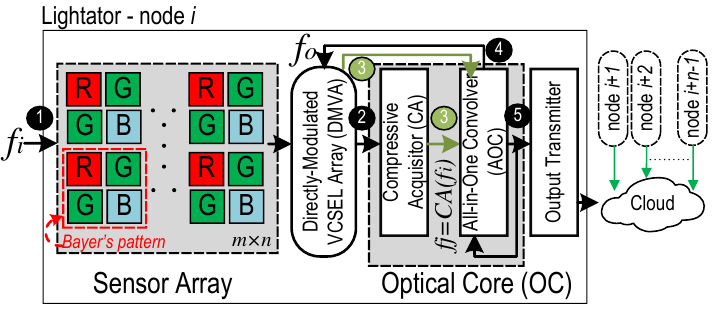}\vspace{-1.4em}
 \end{tabular} 
\caption{High-level operational flow of Lightator.}
\label{overall} \vspace{-1em}
\end{center}
\end{figure}

We propose Lightator as a high-performance, energy-efficient, and versatile PNS accelerator with compressive acquisition for real-time image processing at the edge. The key idea behind developing such an architecture is to have a standalone optical framework (not relying on off-chip processors \cite{xu2020macsen,angizi2023pisa,tabrizchi2023appcip}) for the first time to compress and process all layers in Multi-Layer Perceptron (MLPs) and CNNs in a low-bit-width fashion to tailor the trade-offs between the power consumption and accuracy.  
The high-level operational flow of Lightator represented by node \textit{i} in a multi-node IoT structure is shown in Fig. \ref{overall}. The design consists of a m$\times$n sensor array and an ultra-fast Optical Core (OC) interfacing through a Directly-Modulated VCSEL Array (DMVA). In step \encircle{1}, the input frame $f_i$ is captured by a global-shutter RGB image sensor and processed in an innovative ADC-less fashion with the DMVA unit. In \encircle{2}, the resulting waveguides can be optionally fed to an OC's  Compressive Acquisitor (CA) unit that reduces the spatial dimension by mean pooling across channels and strided convolution to generate $f_j$. This step can be readily skipped depending on the workload and requirements. In \encircle{3}, the All-in-One Convolver (AOC) processes the DNN layer and transmits the results $f_o$ in step \encircle{4} to be used by DMVA as the input to the next layer. Therefore, step \encircle{3}$\leftrightarrow$\encircle{4} features layer-by-layer DNN process through a novel hardware mechanism to reuse DMVA and eliminates the need for conventional area-/power-consuming activation banks \cite{sunny2021crosslight,sunny2021robin} and prepares the result for transmitter \encircle{5}.

\begin{figure}[t]
\begin{center}
\begin{tabular}{c}
\includegraphics [width=0.99\linewidth]{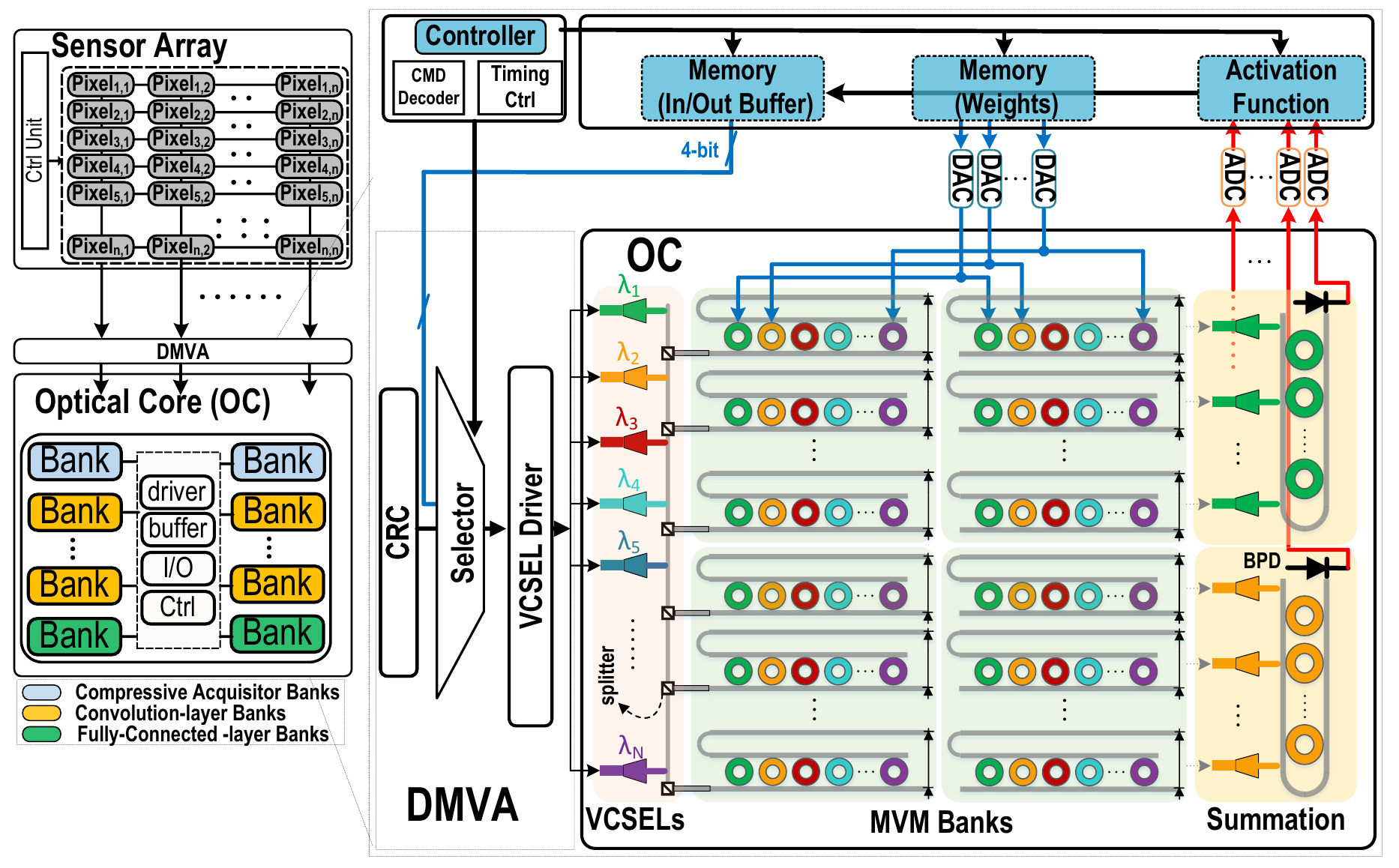}\vspace{-1.4em}
 \end{tabular} 
\caption{Lightator architecture consisting of a sensor array and the optical core.}
\label{arch} \vspace{-2.7em}
\end{center}
\end{figure}

The detailed architecture of Lightator is presented in Fig. \ref{arch}. A sensor array in an ADC-less fashion is connected to the VCSEL driver circuit using a Comparator-based pixel Reading Circuit (CRC). VCSEL driver drives an array of VCSELs in the OC. The Matrix-Vector Multiplication (MVM) banks and the subsequent summation section handle the execution of MAC operations across different network layers. The primary advantage of Lightator processing core is that it only requires mapping weight data onto MRs, while activation values are directly modulated onto the core's input light through VCSELs by adjusting their driving currents, unlike prior designs discussed in the previous section. This configuration allows the entire capacity of the OC to be dedicated to weight values rather than activation, resulting in substantial energy savings, as driving VCSELs consumes significantly less power compared to tuning MRs \cite{sunny2021crosslight,sunny2021robin}. Moreover, additional energy efficiency is achieved as the VCSEL driver directly takes the digital output of the previous layer, eliminating the need for conversion to analog MR tuning signals using DACs. The electronic component on top consists of the activation functions supporting $Sign$, $ReLU$, and $tanh$, as well as the storage for the weights and the activated feature maps from previous layers. A more detailed explanation of the various components of Lightator architecture is provided in the following.

\textbf{ADC-Less Imager.}
A 256$\times$256 global-shutter RGB image sensor has been considered in the presented design. Every pixel's Photo-Diode (PD) generates a photo-current with respect to the external light intensity which in turn leads to a voltage drop ($V_{PD}$).   
By utilizing the CRC, the usage of power-hungry and area-consuming ADCs is resolved.
The CRC is responsible for reading the output of the pixel and so the analog output of pixels will be converted to 4-bit digital data.

\begin{figure}[b]
\begin{center}\vspace{-2em}
\begin{tabular}{ll}
\includegraphics [width=0.43\linewidth]{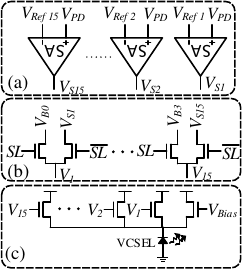} &
\includegraphics [width=0.50\linewidth]{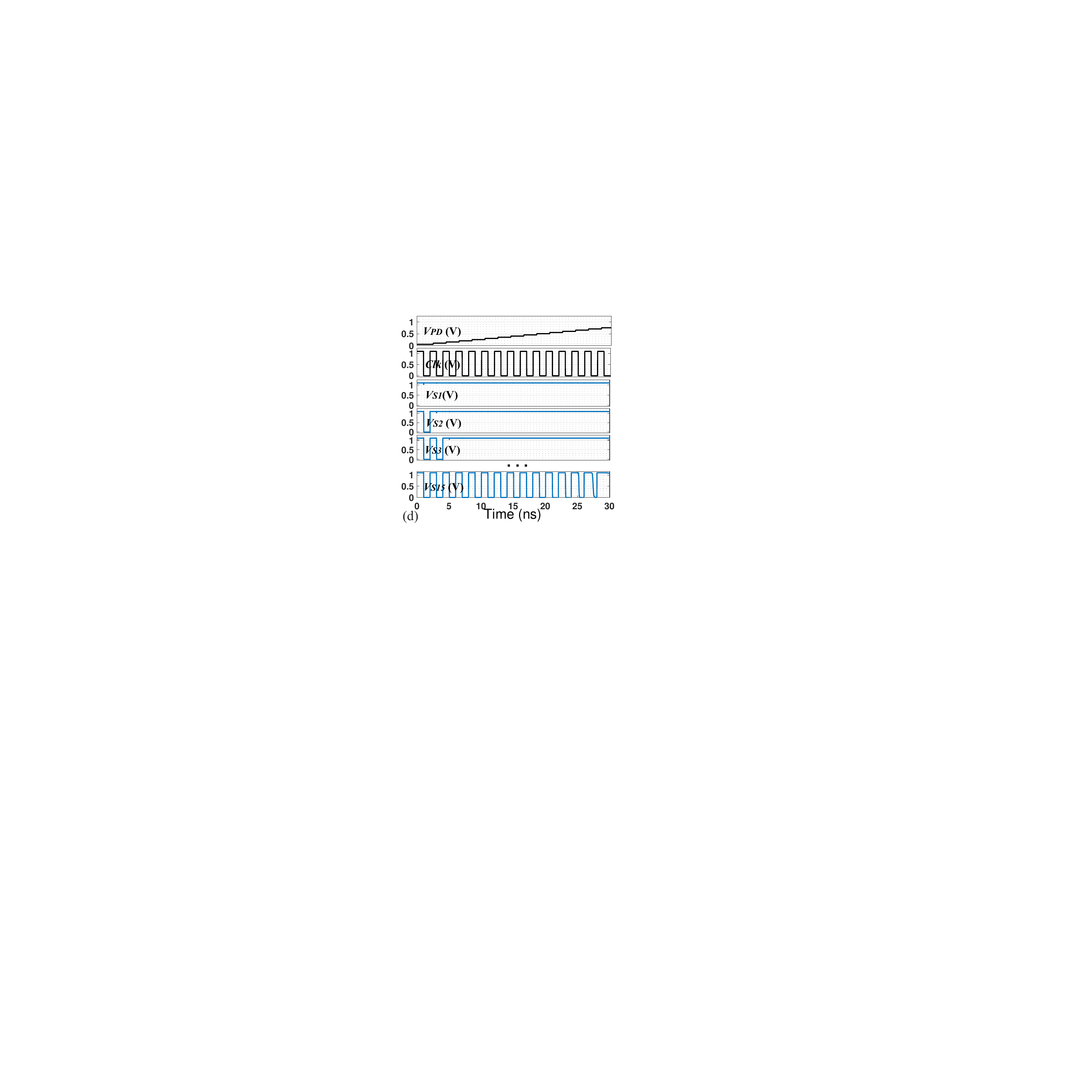}\\
 \end{tabular} \vspace{-1em}
\caption{Components of the DMVA: (a) CRC, (b) Selector, (c) VCSEL driver, (d) Sample waveforms of CRC input from the pixel and respective outputs .} \vspace{-2em}
\label{pri}
\end{center}
\end{figure}

\textbf{Directly-Modulated VCSEL Array.} 
DMVA is developed to convert its electrical input to light with a specific wavelength and intensity. Instead of generating raw light, the intensity of light generated by VCSELs is correlated with the input data of the VCSEL driver, and the wavelength is correlated with the VCSEL structure itself. The input of the VCSEL driver comes from either the pixel array or the output of the previous layer which is processed by the OC. This input is modulated to a specific wavelength and fed to OC as activation to participate in the MAC operation of the next DNN layer. The DMVA consists of three components as shown in Fig. \ref{pri}: CRC, Selector, and VCSEL driver. Each CRC unit (Fig. \ref{pri}(a)) contains 15 voltage comparators and is utilized instead of ADCs to read the pixel's output voltage. CRC receives pixel's $V_{PD}$ and compares it with 15 reference voltages ($V_{Ref}$) which are spanned in the range of pixel output voltage. According to the value of the $V_{PD}$, the output of the comparators ($V_{S}$) will be either `0' or `1' and later these binary voltages will be used to control the VCSEL's driving transistors. Fig. \ref{pri}(d), depicts a sample waveform of the pixel's output voltage and comparator outputs that are used for controlling the driving current of the VCSELs. According to Fig. \ref{pri}(d), by increasing the $V_{PD}$, more number of comparators outputs ($V_{S}$) will be `1' leading larger number of ON transistors in the VCSEL driver circuit. 

A selector circuit is used to select the input-controlling voltages of the VCSEL driver as depicted in Fig. \ref{pri}(b). 
During processing the first layer of the network, the selector connects the output of the pixel array to VCSEL and later when the rest of the layers are getting processed, the selector connects the output of previous network layers as the input of VCESL driver to be modulated and fed as the activation of next layers.
The VCSEL driver circuit (Fig. \ref{pri}(c)) comprises 16 parallel driving transistors that encode 4-bit data. Depending on input signals from either the CRC ($V_{S}$) or the output of the previous layer ($V_{B}$) coming from the selector in Fig. \ref{pri}(b), the number of transistors supplying VCSEL's driving current will be adjusted. When pixel voltage is large or the digital input from the previous layer is greater, more driving transistors will be activated, leading to an increase in the light intensity generated by VCSEL.

\textbf{Optical Core.} OC's MR-based computational units are virtually divided into multiple banks as color-coded in Fig. \ref{arch} including compressive acquisitor, convolutional layer, and fully-connected layer to execute various DNNs all through adjusting weight parameters and MVM operation if required. 

\begin{figure}[t] 
\centering
\includegraphics [width=0.99\linewidth]{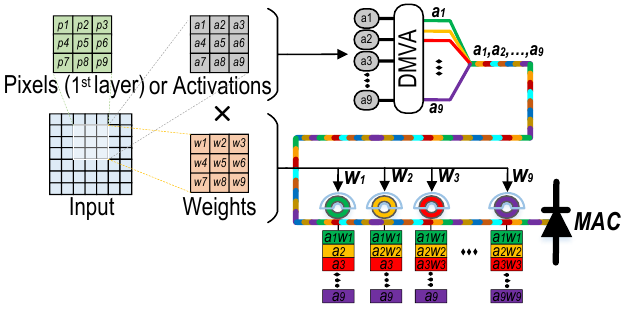}
\vspace{-1em}
\caption{Implementing a 3$\times$3 kernel in an arm.}
 \vspace{-2em}
\label{mapping1}
\end{figure}

\textit{1. All-in-One Convolver (AOC):}
The OC comprises three main components as depicted in Fig. \ref{arch}, VCSELs, MVM banks, and the summation section. VCSELs generate light waves that represent activation values, with the intensity of the light corresponding to these values. MVM banks contain MRs that are mapped with weight values and partitioned in the arms. The MRs adjust the intensity of incoming light based on their mapped weight values, affecting only light with the same wavelength as the MR. This process involves multiplying the activation's light intensity with the weight stored in the MR that is shown in Fig. \ref{mapping1}. To perform MAC operation, a light signal containing all of the required activation values that are modulated on different wavelengths passes through the arm housing MRs with mapped weights. As the light passes the arm, each MR influences the intensity of light at a wavelength corresponding to that specific MR. A Balanced PhotoDetectors (BPD) at the end of each arm handles accumulation, enabling MAC operations to be performed in each arm of the MVM bank. 

The number of multiplication that can be conducted inside an arm is equal to the number of MRs in the arm. In the case of processing fully connected layers or convolutional layers with large kernel sizes that require MAC operation of a large number of activations and weights, the number of multiplications exceeds the capacity of the arm. In these cases, a large number of MACs are divided into smaller segments that can fit within an arm. Subsequently, these segmented MAC results are summed in the summation section to obtain the final MAC result.
To facilitate the processing of multiple layers and enable the processing of the entire neural network on the Lightator platform, in addition to the optical core, an electronic part is required. This part is essential for storing the weight values of various layers since, due to the core's physical limitations, all of the weights of a network cannot be simultaneously mapped to the optical core. Thus, weight values are stored in a dedicated memory and then mapped to the MRs during the processing of each layer. Another memory is utilized to retain the processed output from the network's previous layer, which is subsequently fed as activation to the next layer. In addition, implementing an activation function at the end of each layer is more efficient in the electronic domain than the optic domain \cite{sunny2021crosslight,sunny2021robin}, thus, the electronic part is responsible for performing the activation function. The controller unit controls the procedure and timing of the platform.

\textit{2. Compressive Acquisitor (CA):}
CA banks are dedicated to serving as a compression/pooling layer, where an RGB-to-grayscale conversion and/or configurable average pooling can be done all through adjusting MRs. We propose to conduct the compression in a single operational cycle by mapping proper compression weights to the OC banks and performing the corresponding MAC operation. The conversion from RGB to grayscale can be achieved by forming a weighted sum of the R, G, and B pixel values after CRC as
$P_{Grayscale}=(0.299\times P_{R})+(0.587\times P_{G})+(0.114\times P_{B})$. And, as an example, the 2$\times$2 average pooling layer containing $P_{1}$ to $P_{4}$ pixels can be formulated similar to a weighted sum as follows:
$P_{Avg}=(0.25\times P_{1} )+(0.25\times P_{2})+(0.25\times P_{3})+(0.25\times P_{4})$. Therefore, a nicely-compressed and gray-scale-converted input can be given by properly tuning the weight parameters as follows.
\begin{equation} \label{compression} 
\begin{split} 
   \tiny 
P_{AvgGray}=(0.25\times 0.299\times P_{1R} )+ (0.25 \times 0.587\times P_{1G}) \\ +(0.25 \times 0.114\times P_{1B})+ ...+ (0.25\times 0.299\times P_{4R} )+ \\ (0.25 \times 0.587\times P_{4G})+(0.25 \times 0.114\times P_{4B})
\end{split}
\end{equation} 
Where in $P_{ij}$, \textit{i} is the pixel number identifier and \textit{j} denotes the channel which can be R, G, or, B.  By using the above method and mapping the coefficients of the resultant equation (\ref{compression}) in the OC's MR banks, RGB-to-grayscale conversion and average pooling of any size can be conducted simultaneously. \vspace{-0.5em}

\section{Hardware Mapping}
\textbf{Methodology.} 
The MVM banks are the most crucial parts of the OC. For the processing of the compression layer, convolutional layer, or fully connected layer, the respective weights must be assigned to the MRs within the OC banks. In our design, MRs are organized into groups of 9 inside each arm. It is worth mentioning that due to the widespread use of a kernel size of 3$\times$3 in most CNNs, the number of MRs in each arm is considered as 9 to enable efficient performance of a stride of 3$\times$3 kernel. Then each set of 6 arms is treated as a bank. In total, 96 banks are arranged in an array with 8 columns and 12 rows to form the main processing part of our OC. With each bank comprising 9$\times$6=54 MRs, the MVM banks collectively house 5184 MRs. This implies that, at maximum, 5184 MAC operations can be executed in each operational cycle of the OC. 

\begin{figure}[b]
\begin{center} \vspace{-2em}
\begin{tabular}{c}
\includegraphics [width=1\linewidth]{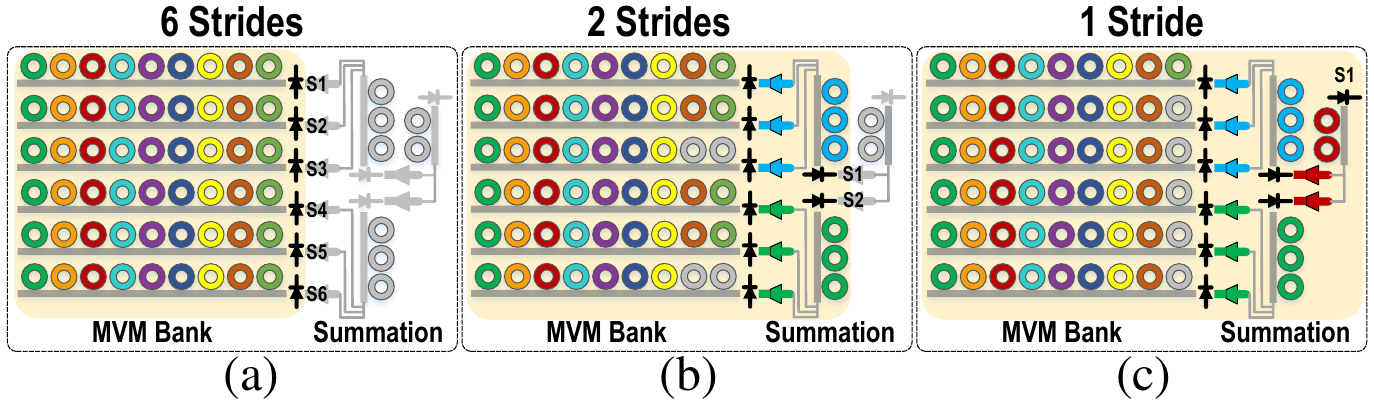}\vspace{-1.4em}
 \end{tabular} 
\caption{Hardware mapping for (a) 6 Strides (3$\times$3), (b) 2 Strides (5$\times$5), (c) 1 Stride (7$\times$7).}
\label{mapping2} \vspace{-1.1em}
\end{center}
\end{figure}

Fig. \ref{mapping2} illustrates the bank configuration engaged in MAC operations within a convolutional layer, utilizing kernel sizes of 3$\times$3, 5$\times$5, and 7$\times$7, with their respective summation sections positioned at the right end of each bank. As the configuration of OC is specified for 3$\times$3 kernels, as depicted in Fig. \ref{mapping2}(a), all of the MRs are allocated to be mapped with weight values enabling each arm to execute a stride. BPD performs the summation operation and the MAC result can be directly sent out without utilizing the summation component. As a result, the summation component will be inactive and inoperative (depicted in gray in Fig. \ref{mapping2}(a)). Under these circumstances, each bank can execute 6 strides of a convolution operation. When dealing with a 5$\times$5 kernel, we have 25 weight values that need to be mapped on MRs. Thus, 3 arms of the bank are allocated for one stride. As 27 MRs are available in 3 arms, in each set of 3 arms, 2 MRs remain unused and inactive, indicated by the gray shading in Fig. \ref{mapping2}(b). Since each arm lacks the capability to sum the 25 multiplication elements, an additional summation of the partial sum results becomes necessary. The initial stage of the summation part, depicted in Fig. \ref{mapping2}(b), is responsible for doing that, while the second stage, represented in gray color, remains inactive and unused. As illustrated in Fig. \ref{mapping2}(b), in the case of 5$\times$5 kernel size, each bank can perform 2 strides. For a 7$\times$7 kernel size, a total of 49 MRs are necessary for weight mapping, leading to the entire bank being dedicated to a single stride.
Nevertheless, 5 MRs per bank remain inactive and unused, shown in gray in Fig. \ref{mapping2}(c). Through further processing involving two stages of the summation part, partial products are combined, allowing the final MAC results to be sent out. In the case of fully-connected layers, we segment the entire MAC operations into sets of 9 MACs, map their corresponding weights to arms, and subsequently aggregate the partial results using the summation part to derive the ultimate MAC result.

\vspace{-0.7em}

\section{Experiments}
\begin{figure}[t] 
\centering
\includegraphics [width=0.9\linewidth]{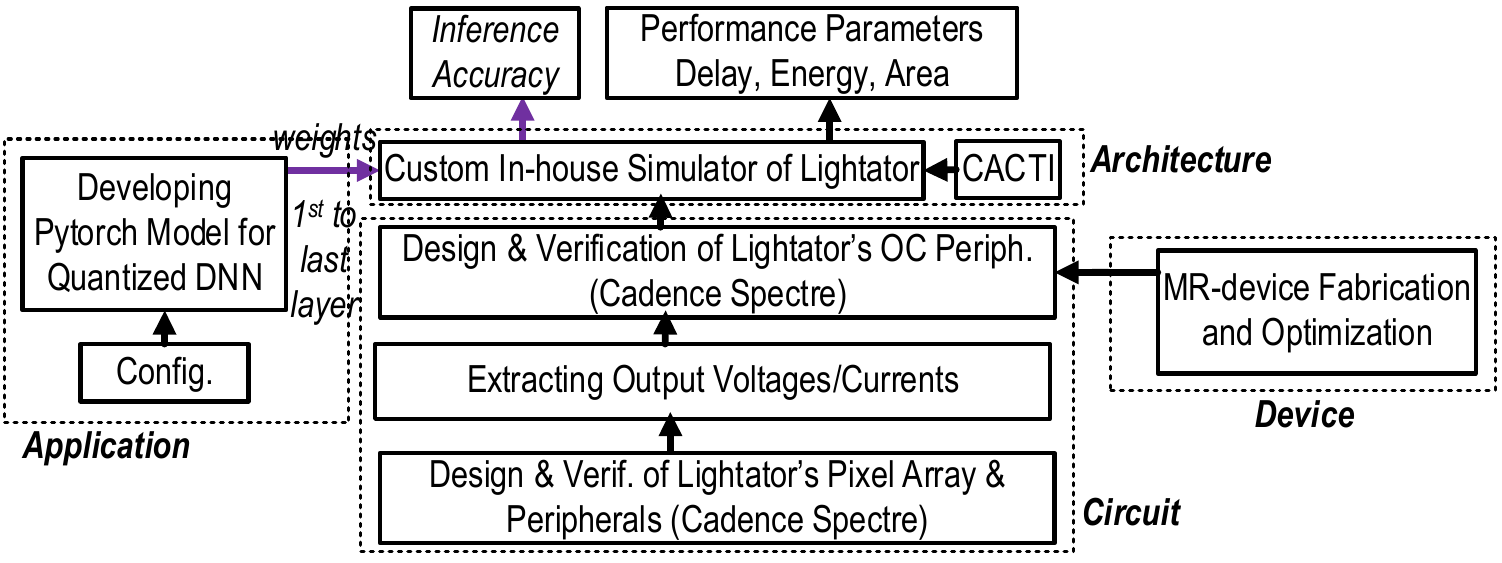}
\vspace{-1.1em}
\caption{Proposed bottom-up evaluation framework.}
\vspace{-2em}
\label{framework}
\end{figure}

\textbf{Framework.} 
As shown in Fig. \ref{framework}, the assessment framework consists of device-, circuit-, architecture-, and application-level components. At the device level, we manufactured and fine-tuned the MR devices and obtained the circuit parameters for co-simulation with interface CMOS circuits in Cadence Spectre and SPICE. Progressing to the circuit level, we initially implement the pixel's array and peripheral circuitry using the 45nm NCSU Product Development Kit (PDK) library \cite{NCSU_PDK} in Cadence, from which we derive the output voltages and currents. Then we proceed to develop all Lightator's components excluding kernel banks (implemented in Cacti \cite{thoziyoor2008cacti}) in Cadence Spectre. At the application level, we train PyTorch models w.r.t. the under-test DNN models and datasets and extract weight parameters. These parameters are then quantized and mapped into the OC for adjusting MR elements. To preserve optimal accuracy post-precision reduction, we undertake an additional six epochs of training employing quantization-aware techniques. This ensures the model's robustness and performance integrity in the face of reduced precision.
At the architecture level, we develop a custom in-house simulator for Lightator to work with the 1$^{st}$-to-last layer weight parameters and calculate both the execution time and power consumption required for the DNN models as well as inference accuracy. Moreover, it offers flexibility in terms of MVM array configuration and the selection of peripheral designs. We conduct experiments on Lightator considering various [Weight: Activation] configurations with several datasets, including MNIST evaluated on LeNET, and CIFAR10, and CIFAR100 on VGG9.

\begin{figure}[b] \vspace{-1em}
\centering
\includegraphics [width=0.82\linewidth]{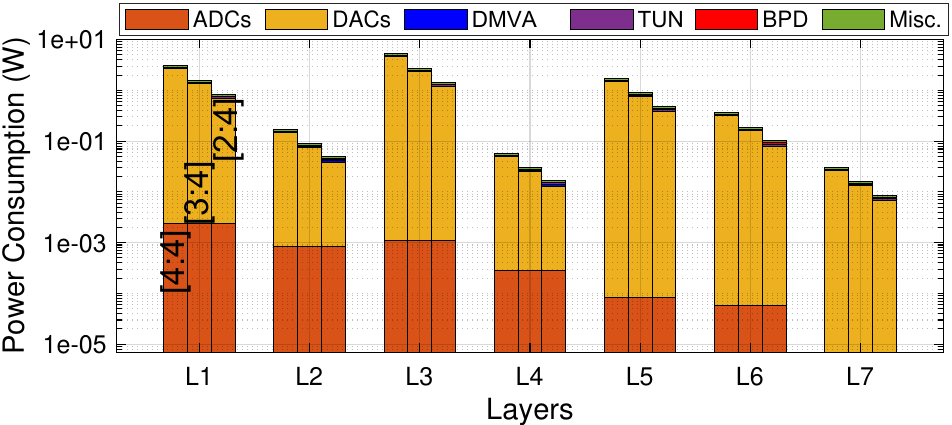}
\vspace{-1.3em}
\caption{Break-down of power consumption for LeNET on [4:4], [3:4], and [2:4]. Note: Pooling layers are implemented within CA banks with pre-set weight coefficients.}
\vspace{-1em}
\label{lenetpower}
\end{figure}

\begin{figure}[t] 
\centering
\includegraphics [width=0.86\linewidth]{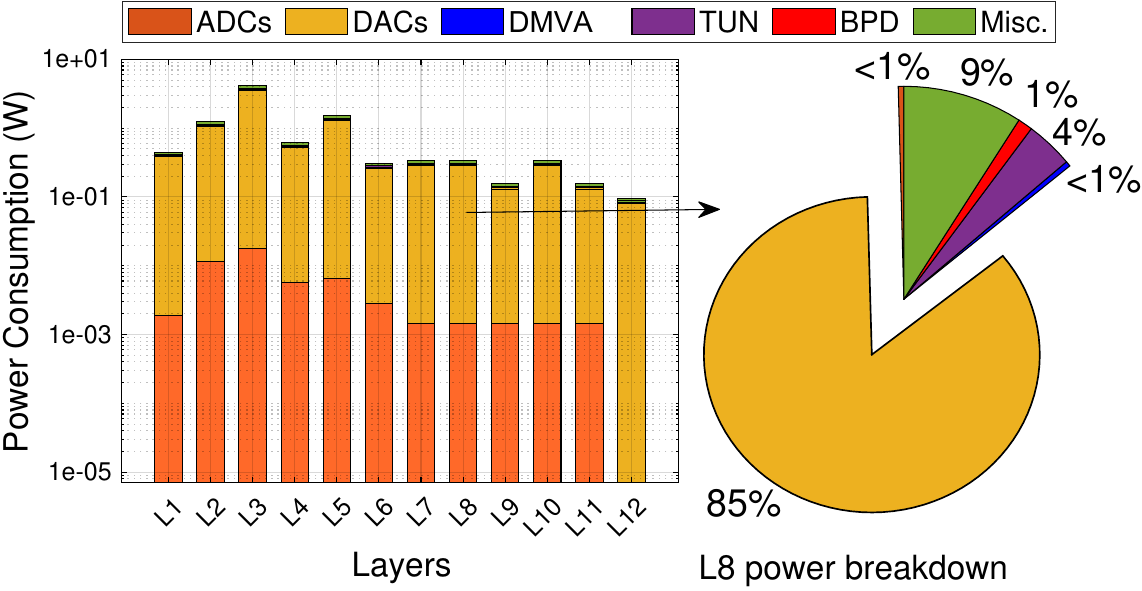}
\vspace{-1.3em}
\caption{Break-down of power consumption for VGG9 on [3:4] configuration.}
\vspace{-1.4em}
\label{vggpower}
\end{figure}

\textbf{Power Consumption \& Performance.} 
Fig. \ref{lenetpower} shows the layer-wise breakdown of components of power consumption including ADCs, DACs, DMVA (with CRC, VCSELs, and drivers), Tuning circuitry (TUN), BPDs, and Misc. (Controller, etc.) for LeNET model mapped to Lightator. Lightator effortlessly implements all convolutional and pooling layers (indicated by $L_i$) for three weight and activation [W:A] configurations of [4:4], [3:4], and [2:4]. Pooling layers are implemented within CA banks with pre-set weight coefficients. We observe that decreasing the bit-width of weight parameters for each layer results in power saving for the edge device, where on average 2.4$\times$ more power efficiency is reported. 
This mainly comes from power-gating parts of the 4-bit DAC circuits that are related to its extra bit precision, when they process 3-bit and 2-bit data.
In Fig. \ref{vggpower}, the distribution of power consumption components for the VGG9 model is depicted layer-by-layer, specifically focusing on configurations limited to [3:4]. We leverage CA banks for a light compression of input images as the proof-of-concept before feeding them into the model. This leads to a 42.2\% reduction in power consumption of the first layer. The pie chart in Fig. \ref{vggpower} clarifies the breakdown of power consumption in a sample layer as well. We observe that consistently across all layers, DACs contribute to more than 85\% of the total power consumption, as DAC usage is required to convert all of the weight values to analog inputs for tuning purposes.

\textbf{Comparison with Optical Accelerators.} 
Table \ref{results} provides our comprehensive simulation results for selected MR-based optical accelerators and Lightator in various [W:A] configurations compared with the baseline, an NVIDIA Geforce RTX 3060Ti GPU. The under-test DNN accelerators includes LightBulb \cite{zokaee2020lightbulb}, HolyLight \cite{liu2019holylight}, HQNNA \cite{sunny2022silicon}, Robin \cite{sunny2021robin}, and CrossLight \cite{sunny2021crosslight} discussed in the background section. To ensure an unbiased assessment, we created the designs from the ground up resembling the original design, employing the evaluation framework and our in-house simulator, and reported the results in a reasonable area constraint for all accelerators ($\sim$20-60$mm^2$). Our framework features 96 banks, each comprising 6 arms with 9 MRs. \vspace{-1em}

\begin{table}[h]
\caption{Performance comparison with optical designs.}\vspace{-1em}
\scalebox{0.6}{
\begin{tabular}{lcccccc}
\hline
\rowcolor[HTML]{C0C0C0} 
\multicolumn{1}{c}{\cellcolor[HTML]{C0C0C0}}                                                                                           & \cellcolor[HTML]{C0C0C0}                                                                                       & \cellcolor[HTML]{C0C0C0}                                                                                    & \cellcolor[HTML]{C0C0C0}                                  & \multicolumn{3}{c}{\cellcolor[HTML]{C0C0C0}\textbf{Accuracy (\%)}}                                \\ \cline{5-7} 
\rowcolor[HTML]{C0C0C0} 
\multicolumn{1}{c}{\multirow{-2}{*}{\cellcolor[HTML]{C0C0C0}\textbf{\begin{tabular}[c]{@{}c@{}}Designs \&\\ $[$W: A$]$\end{tabular}}}} & \multirow{-2}{*}{\cellcolor[HTML]{C0C0C0}\textbf{\begin{tabular}[c]{@{}c@{}}Process node\\ (nm)\end{tabular}}} & \multirow{-2}{*}{\cellcolor[HTML]{C0C0C0}\textbf{\begin{tabular}[c]{@{}c@{}}Max Power \\ (W)\end{tabular}}} & \multirow{-2}{*}{\cellcolor[HTML]{C0C0C0}\textbf{KFPS/W}} & \multicolumn{1}{l}{\cellcolor[HTML]{C0C0C0}\textbf{MNIST}} & \textbf{CIFAR10} & \textbf{CIFAR100} \\ \hline
baseline [32:32]$^\S$                                                                                                                  & 8                                                                                                              & 200                                                                                                         & -                                                         & 98.53                                                      & 90.46            & 67.8              \\
LightBulb [1:1] \cite{zokaee2020lightbulb}                                                                                             & 32                                                                                                             & 68.3                                                                                                        & 57.75                                                     & 96.7                                                       & -                & -                 \\
HolyLight [4:4] \cite{liu2019holylight}                                                                                                & 32                                                                                                             & 66.9                                                                                                        & 3.3                                                       & 98.9                                                       & 88.5             & -                 \\
HQNNA \cite{sunny2022silicon}                                                                                                          & 45                                                                                                             & -                                                                                                           & 34.6                                                      & -                                                          & 89.68            & 61.95             \\
Robin [1:4] \cite{sunny2021robin}                                                                                                      & 45                                                                                                             & 106                                                                                                         & 46.5                                                      & -                                                          & 62.5             & 45.6              \\
CrossLight [4:4] \cite{sunny2021crosslight}                                                                                            & -$^{*}$                                                                                                        & 84-390                                                                                                      & 10.78-52.59                                               & 92.6                                                       & 78.85            & -                 \\
\textbf{Lightator [4:4]}                                                                                                               & \textbf{45}                                                                                                    & \textbf{5.28}                                                                                               & \textbf{61.61}                                            & \textbf{98.12}                                             & \textbf{88.87}   & \textbf{64.22}    \\
\textbf{Lightator [3:4]}                                                                                                               & \textbf{45}                                                                                                    & \textbf{2.71}                                                                                               & \textbf{117.65}                                           & \textbf{98.05}                                             & \textbf{86.3}    & \textbf{61.04}    \\
\textbf{Lightator [2:4]}                                                                                                               & \textbf{45}                                                                                                    & \textbf{1.46}                                                                                               & \textbf{188.24}                                           & \textbf{93.95}                                             & \textbf{70.55}   & \textbf{41.4}     \\
\textbf{Lightator-MX [4:4][3:4]$^{\dagger}$}                                                                                           & \textbf{45}                                                                                                    & \textbf{3.64}                                                                                               & \textbf{84.4}                                             & \textbf{97.85}                                             & \textbf{85.65}   & \textbf{63.37}    \\
\textbf{Lightator-MX [4:4][2:4]$^{\ddagger}$}                                                                                          & \textbf{45}                                                                                                    & \textbf{1.97}                                                                                               & \textbf{126.6}                                            & \textbf{94.8}                                              & \textbf{78.87}   & \textbf{51.29}    \\ \hline
\end{tabular}}

\tiny$^{\S}$NVIDIA Geforce RTX 3060Ti GPU.
$^{*}$Data is not reported/not achievable in the paper. $^{\dagger}$Lightator with mixed-precision scheme, where L1[4:4] - L2:LN [3:4]. $^{\ddagger}$Lightator with mixed-precision scheme, where L1[4:4] - L2:LN [2:4]. \vspace{-1.4em}
\label{results}
\end{table}

Here we list our key observations. 
$(1)$ We observe Lightators's variants demonstrate remarkable power efficiency over counterpart designs on the VGG9 model running CIFAR100, e.g., Lightator [3:4] consumes 2.71 W which can be drawn from the low power budget of edge devices, however, the best low-power accelerator, i.e., HolyLight \cite{liu2019holylight} requires 66.9 W or higher \cite{sunny2021robin}. Such striking power efficiency comes from $(i)$ eliminating the MRs tuned by activation parameters which results in saving the tuning power required for the MRs, and $(ii)$ reducing the additional power and area requirements caused by the extensive utilization of ADCs and DACs. 
$(2)$ On average the Lighator reduces power consumption by $\sim$73$\times$, 24.68$\times$, 30.9$\times$ compared with the baseline [32:32],  HolyLight [4:4] \cite{liu2019holylight}, and CrossLight [4:4] \cite{sunny2021crosslight}, respectively.
$(3)$ As we reduce the weight bit-width, the power consumption can be reduced at the cost of accuracy degradation, where Lightator [3:4] achieves $\sim$2$\times$ power saving at the cost of 3.17\% accuracy drop.
$(4)$ The mixed precision implementation of DNNs on Lightator is termed as Lightator-MX in Table \ref{results} in which the first layer configuration is kept to [4:4] and the rest of the layers are processed in [3:4] or [2:4] precision. We observe the trade-offs between power consumption and accuracy that can be readily adjusted based on the image-processing task requirements. As shown, Lightator-MX [4:4][3:4] as an optimal design imposes $\sim$0.9 W extra power consumption to the Lighatator [3:4] increasing the accuracy of CIFAR100 by 2.33\%. $(5)$ As for the throughput ($\frac{frame}{second}$) per watt, Lightator [3:4] demonstrates 117.65 kilo FPS/W increasing inference performance by $\sim$2$\times$ compared to the best result reported for LightBulb \cite{zokaee2020lightbulb}. Overall, considering the test accuracy results over three under-test data-sets, Lightator-MX [4:4][3:4] offers the best performance-quality number with 84.4 kilo FPS/W. 
$(6)$ As for only accuracy, our experiments generally reveal that Lightator with [3:4] and [4:4] configurations could demonstrate acceptable accuracy over three under-test data-sets. On MNIST and CIFAR10 data-sets, Lightator [4:4] achieves the second-highest accuracy among all optical accelerators after HolyLight \cite{liu2019holylight} and HQNNA \cite{sunny2022silicon}, respectively, while showing higher KFPS/W compared to them. 
$(7)$ We observe that activations and weights exhibit increasing sensitivity to changes in bit-width. 

\begin{figure}[t] 
\centering
\includegraphics [width=0.80\linewidth]{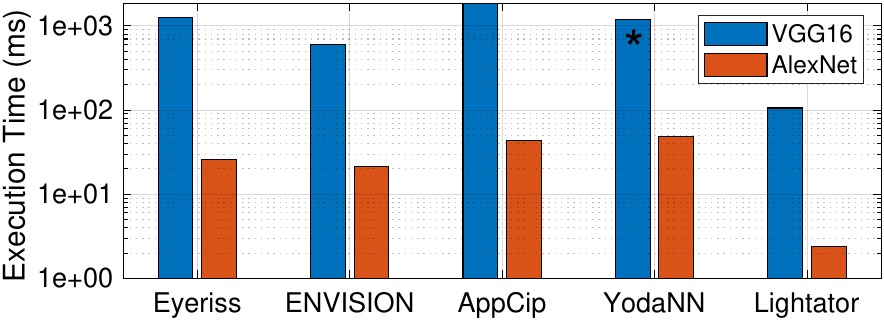}
\vspace{-1.3em}
\caption{Log-scaled execution time of various accelerators. YodaNN's results for VGG16 are substituted with VGG13.}
\vspace{-2em}
\label{exe}
\end{figure}

\textbf{Comparison with Electronic Accelerators.} 
To demonstrate the intrinsic parallelism observed in Lightator as an optical accelerator, we further explore its execution time compared with four well-known digital electronic accelerators, each with a distinct parallelism technique and hardware mapping method, i.e., Eyeriss \cite{chen2017eyeriss}, YodaNN \cite{andri2018yodann}, AppCip \cite{tabrizchi2023appcip}, and ENVISION \cite{moons201714} running VGG16 and AlexNet. Eyeriss employs a spatial architecture that utilizes row-stationary dataflow to minimize energy consumption. YodaNN is an ASIC accelerator optimized for binary-weight CNNs with support for different filter sizes in parallel. AppCip as a PIS implements
instant RGB-to-grayscale conversion, highly parallel analog convolution-in-pixel, and low-precision quinary weight neural networks. ENVISION utilizes subword parallel MACs with dynamic adjustments to voltage, frequency, and bit precision scaling. The simulation results plotted in Fig. \ref{exe} demonstrate the superiority of the optical accelerator in processing DNN layers compared with the electronic ones over both models. We observe Lightator reduces the execution time by a factor of 10.7$\times$, 20.4$\times$, 18.1$\times$, 8.8$\times$ over Eyeriss \cite{chen2017eyeriss}, YodaNN \cite{andri2018yodann}, AppCip \cite{tabrizchi2023appcip}, and ENVISION \cite{moons201714} on AlexNet, respectively. A similar trend is observable for the VGG16 model. \vspace{-1em} 

\section{Conclusion}\vspace{-0.5em} 
Here, we presented an efficient optical near-sensor accelerator for vision applications named Lightator. Our design features innovative compressive acquisition of input frames and fine-grained convolution operations for low-power and versatile image processing at the edge. Our results demonstrate that with acceptable accuracy, Lightator achieves 84.4 Kilo FPS/W and reduces power consumption by a factor of $\sim$24$\times$ and 73$\times$ on average compared with recent photonic accelerators and GPU baseline.\vspace{-1.1em}

\section*{Acknowledgment}\vspace{-0.5em}
\small This work is supported in part by the National Science Foundation (NSF) under grant no. 2228028, 2216772, 2046226, 2006788, 2216773, 2303114, 2127780, 2319198, 2321840, 2312517, 2235472, Semiconductor Research Corporation (SRC), ONR Young Investigator Program Award, ONR \#N00014-21-1-2225 and \#N00014-22-1-2067, and the Air Force Office of Scientific Research under award \#FA9550-22-1-0253. 

\vspace{-1em}

\bibliographystyle{ACM-Reference-Format}
\bibliography{Reference}\vspace{-2em}

\end{document}